Available online at www.sciencedirect.com

ScienceDirect

Journal of Electrical Systems and Information Technology 5 (2018) 745–758

www.elsevier.com/locate/jesit# Microcontroller based bidirectional buck–boost converter for photo-voltaic power plant

Viswanatha V. [a,*], Venkata Siva Reddy R. [b]

[a] *Department of Electrical and Electronics Engineering, REVA University, Karnataka 560064, India*
[b] *Department of Electronics and Communication Engineering, REVA University, Karnataka 560064, India*Received 6 May 2016; received in revised form 16 December 2016; accepted 4 April 2017
Available online 27 June 2017## Abstract

A common configuration for a stand-alone PV power system may consist of three converters: a buck converter for the PV panel to charge the battery, a boost converter for the battery to discharge to the load and one for the load voltage regulation. Such a system requires a coordinated control scheme for three converters which can be complicated. A simple structure for a stand-alone PV plant consists of a PV array, a battery unit, and its associated bidirectional converter which is a combination of a buck and boost converter. When controlled properly the system can provide uninterrupted power to the load, despite the intermittent availability of sunlight. In this paper complete design of the converter is carried out and the simulation has been performed using Psim. From the simulation, the graphs are presented to show the converter working in buck mode and boost mode. Controller is designed to take care of mode transition, buck to boost and boost to buck mode automatically based on source voltage. Hardware implementation has been done using microcontroller (8051).
© 2017 Electronics Research Institute (ERI). Production and hosting by Elsevier B.V. This is an open access article under the CC BY-NC-ND license (http://creativecommons.org/licenses/by-nc-nd/4.0/).

*Keywords:* Battery; Microcontroller; DC to DC converter; Solar panel; Battery control strategy## 1. Introduction

In particular, the development of the bidirectional converter as a power interface between main and auxiliary energy storage elements is a key aspect to commercializing photo-voltaic power plant and the advanced electric vehicles such as Hybrid Electric Vehicles (HEVs) and Fuel Cell Electric Vehicles (FCEVs). Not only in the field of automotive industries, it has so many applications like, for interfacing an energy storage device in an autonomous power system like renewable energy generating systems, for super capacitor based energy buffer for electrical gen-sets, for load

---

* Corresponding author.
*E-mail addresses:* vishwanathv@revainstitution.org, viswas779@gmail.com (V. V.), vsreddy@revainstitution.org (V.S.R. R.).Peer review under the responsibility of Electronics Research Institute (ERI).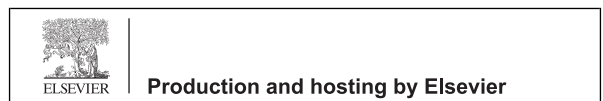

https://doi.org/10.1016/j.jesit.2017.04.0022314-7172/© 2017 Electronics Research Institute (ERI). Production and hosting by Elsevier B.V. This is an open access article under the CC BY-NC-ND license (http://creativecommons.org/licenses/by-nc-nd/4.0/).

leveling systems, DC uninterruptible power supplies, battery chargers, telecom supplies, computer power systems and avionics space systems etc. Stand-alone photovoltaic (PV) power generators are attractive sources of electricity in remote or mountainous locations where a grid connection is not available. A common configuration for a stand-alone PV power system may consist of three converters: a buck converter for the PV panel to charge the battery, a boost converter for the battery to discharge to the load and one for the load voltage regulation. Such a system requires a coordinated control scheme for three converters which can be complicated. In addition the cost of the converters can dominate that of the total system. A simple structure for a stand-alone PV plant consists of a PV array, a battery unit, and its associated bidirectional converter which is a combination of a buck and boost converter. The key devices for efficient operation are the bidirectional buck–boost converter using microcontroller and the battery unit. In Fig. 1 the block diagram of the bidirectional converter is shown.

## 2. Analysis of the proposed bidirectional buck–boost converter

### 2.1. Circuit description

The circuit diagram of the proposed bidirectional converter (BDC) is given in Fig. 1 The proposed BDC consists of a DC bus capacitor $C_{bus}$, two power MOSFETs $S_1$ and $S_2$, a filtering inductor $L_p$, a output capacitor Co and a battery bank. Diodes $D_1$ and $D_2$ are the anti-parallel body diodes of power MOSFETs $S_1$ and $S_2$. If the energy provided from the power source is sufficient, the circuit will operate in the charging mode to store the excess energy into the battery bank. Otherwise, the circuit will operate in the discharging mode to provide the required energy to the load. Thus, the proposed RBC can provide stable and reliable power to the load whether the power source works well. The detailed working processes in the charging, discharging and trickle charging modes are sequentially described as follows.

### 2.2. Charging mode

The equivalent circuit diagrams of the proposed BDC in charging mode are shown in Fig. 2(a) and (b). In charging mode, the proposed BDC works as a buck converter. Thus, the power flow direction of the proposed RBC is from the DC bus to the battery bank to charge the battery with current $I_{batt}$. Waveforms of this mode are also shown in below.

This mode of operation is divided into two states and described as follows:

(a) State 1: In State1 the power MOSFET S1 is turned on, the power MOSFET S2 is turned off and the body diode D2 is reversed as shown in Fig. 2. In this state, the filtering inductor $L_p$ is charged up linearly by the voltage

$$V_L = V_{bus} - V_{batt} \qquad (1)$$

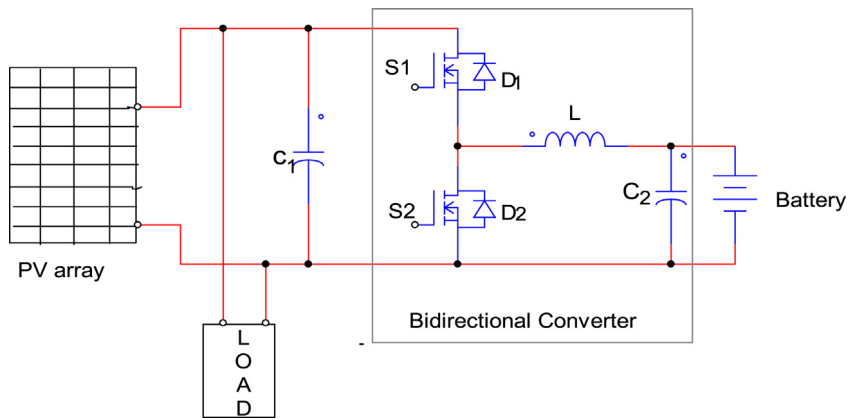

Fig. 1. Circuit diagram of bidirectional converter.

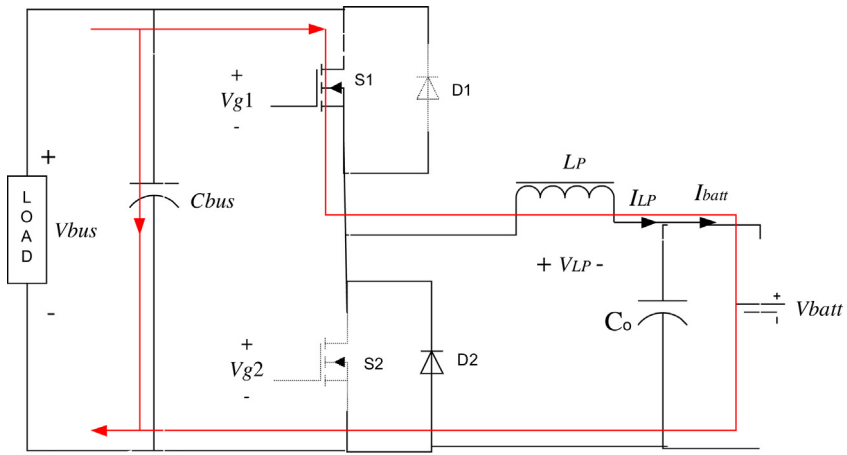

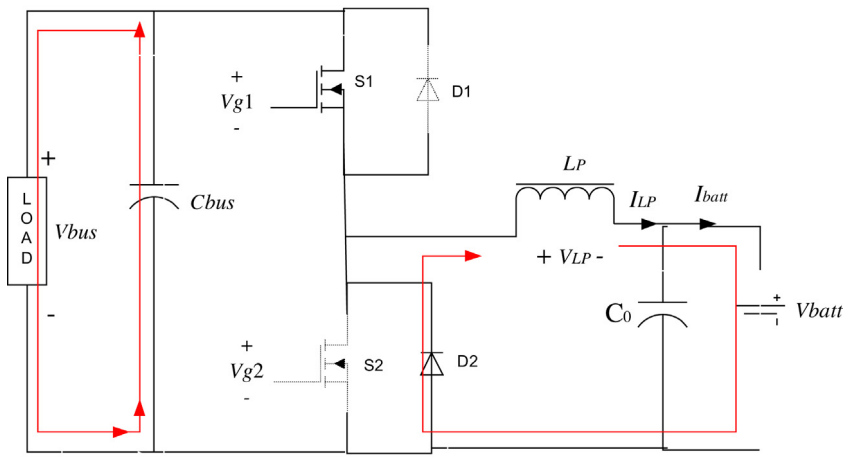

Fig. 2. Charging mode converter.

so that the current in filtering inductor $L_p$ begins to increase linearly with slope

$$I_L = (V_{bus} - V_{batt})/L_p \tag{2}$$

(b) **State 2**: In State2 the power MOSFETs S1 and S2 are turned off and the body diode D2 starts conducting, as shown in Fig. 2(b). In this state, the voltage across filtering inductor $L_p$ is about

$$V_L = -V_{batt} \tag{3}$$

So that the current in filtering inductor $L_p$ begins to decrease linearly with slope

$$I_L = -V_{batt}/L_p \tag{4}$$

The filtering inductor current $I_{Lp}$ (i.e, battery current $I_{batt}$) can be expressed as

$$I_{Lp} = I_{batt} = I_{pp}^* \pm \frac{1}{2}\Delta I_{Lp1} \tag{5}$$

where $I^*_{pp}$ is the desired charging current as shown in waveform and $\Delta L_{p1}$ is the ripple current. The ripple current $\Delta L_{p1}$ can be depicted as

$$\Delta I_{Lp1} = \frac{V_{batt}.(1-D)}{L_p} = \frac{(V_{bus} - V_{batt}).D}{L_p} \tag{6}$$

### 2.3. Discharging mode

In this mode, the proposed BDC works as a boost converter with the current control function. Thus, the power flow direction of the proposed BDC is from the battery bank to the DC bus to discharge the battery with current $I_{batt}$. Specially, the discharge energy of the battery bank is recovered to the DC bus capacity $C_{bus}$ to avoid unnecessary dissipation. Waveforms of this mode are also shown in Fig. 3.

This mode of operation is divided into two states and described as follows:

(a) **state1**: In this state the power MOSFET S2 is turned on and the power MOSFET S1 is turned off, the current in filtering inductor $L_p$ decrease to zero and then charges polarity as shown in Fig. 3(c). In this state, the voltage across filtering inductor $L_p$ is about,

$$V_L = -V_{batt} \tag{7}$$

So that the current in filtering inductor $L_p$ begins to increase linearly with slope

$$I_L = -V_{batt}/L_p \tag{8}$$

(b) **State2**: In this state the power MOSFETs S1 and S2 are turned off and the body diode D1 starts conducting as shown in Fig. 3(d). In this state, the voltage across filtering inductor $L_p$ is about

$$V_L = V_{bus} - V_{batt} \tag{9}$$

So that the current in filtering inductor $L_p$ begins to decrease linearly with slope

$$I_L = (V_{bus} - V_{batt})/L_p \tag{10}$$

Filtering inductor $L_p$ is now discharge. DC bus capacitor $C_{bus}$ starts to be charged by the battery bank discharging current. The filtering inductor current $I_{Lp}$ (i.e, battery current $I_{batt}$) in the state can be expressed as

$$I_{Lp} = I_{batt} = I_{np}^* \pm \frac{1}{2}\Delta I_{Lp2} \tag{11}$$

where $I^*_{np}$ is the desired charging current in the state as shown in Fig. 3(d) and $\Delta I_{Lp2}$ is the ripple current. The ripple current $\Delta I_{Lp2}$ can be depicted as

$$\Delta I_{Lp2} = \frac{V_{batt}.D}{L_p} = \frac{(V_{bus} - V_{batt}).(1-D)}{L_p} \tag{12}$$

### 2.4. Trickle charging mode

This mode is also called as Resting period. In this mode the power MOSFETs S1 and S2 are turned off to disconnect the DC bus and battery bank as shown in Fig. 4. this mode is used to diffuse and distribute the electrolyte ions to neutralize the electrolyte density and increase the battery charging efficiency and life cycle.

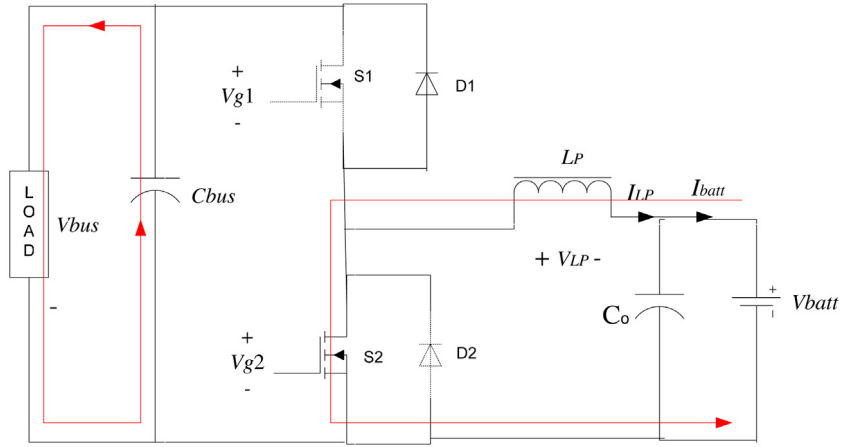

**(c)**

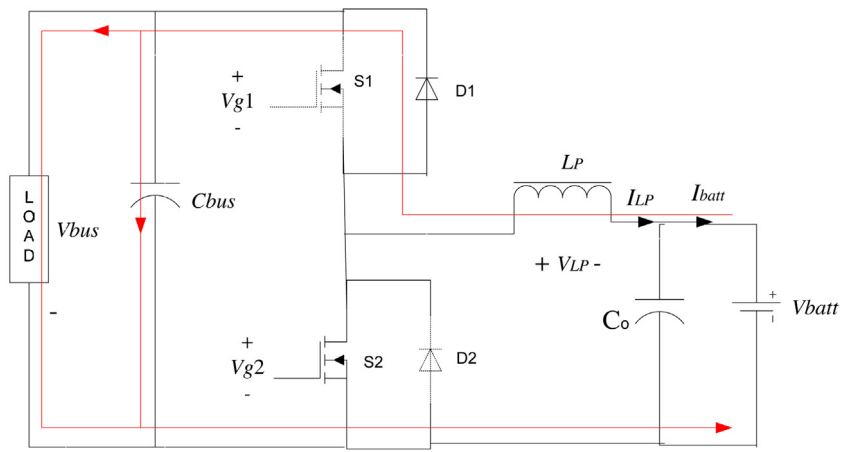

**(d)**

Fig 3. Discharging mode converter.

## 3. Converter design

*3.1. Specifications*

PV array voltage (VP) = 24 V.
PV array current (IP) = 3 A.
Battery voltage (VB) = 12 V.
Switching frequency (Fs) = 20 kHz.
Load voltage (VL) = 24 V.
Load current (IL) = 2.4 A.

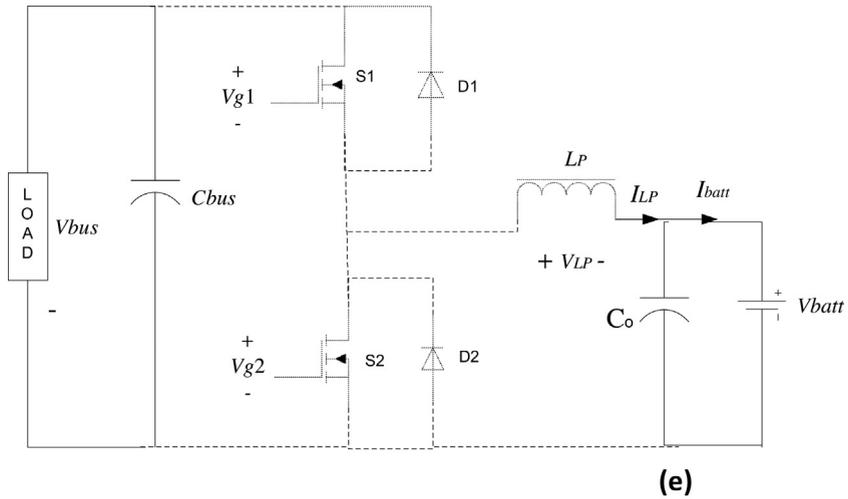

Fig. 4. Trickle charging mode converter.

For Buck Operation:      For Boost Operation:

$$\frac{V_0}{V_S} = D1 \quad : \quad \frac{V_0}{V_S} = \frac{1}{1-D2}$$

$$\frac{V_B}{V_P} = D1 \quad : \quad \frac{V_L}{V_B} = \frac{1}{1-D2}$$

$$D1 = \frac{12}{24} \quad : \quad D2 = \frac{V_L - V_B}{V_L}$$

$$D1 = 0.5 \quad : \quad D2 = 0.5$$

Inductor Design:

$$L_{min} = \frac{V_P * D1 (1-D1)}{F_S * \Delta I} \quad : \quad L_{max} = \frac{V_B * D2}{F_S * \Delta I}$$

$$L_{min} = \frac{24 * 0.5 (0.5)}{20 * 10^3 * 0.3} \quad : \quad L_{max} = \frac{12 * 0.5}{20 * 10^3 * 0.3}$$

$$L_{min} = 1000uH \quad : \quad L_{min} = 1000uH$$

Output Capacitor Design:

Ripple voltage (dv) = 1% of output voltage
    ie 1% of 24v
        dv = 0.24v

$$I_0 = C \frac{dv}{dt}$$

$$C = I_0 \frac{dt}{dv}$$

$$C = \frac{2.4 * 50 * 10^{-6} * 0.5}{0.24}$$

$$C = 250uF$$

## 4. Control technique for bidirectional converter

### 4.1. Voltage mode control

In this technique, voltage across the load is sensed to regulate output voltage. Sensing of voltage using voltage sensor where load voltage is sampled first then it will be sensed by voltage sensor and then its output is given to comparator circuit which compare load voltage then it will take appropriate action to trigger either S1 or s2 based on output voltage. If load voltage is less than fixed or voltage which is to be regulated then battery will supply power to the load through the bidirectional converter. Also using this voltage mode, charging as well as discharging of battery is regulated.

## 5. Simulation and results

PSIM8.0.4 was used for simulating the converter-battery system. Simulation is used for the bi-directional characteristics of the converter. The input voltage at 24 V DC. dc–dc converter is bi-directional in nature (Fig. 5).

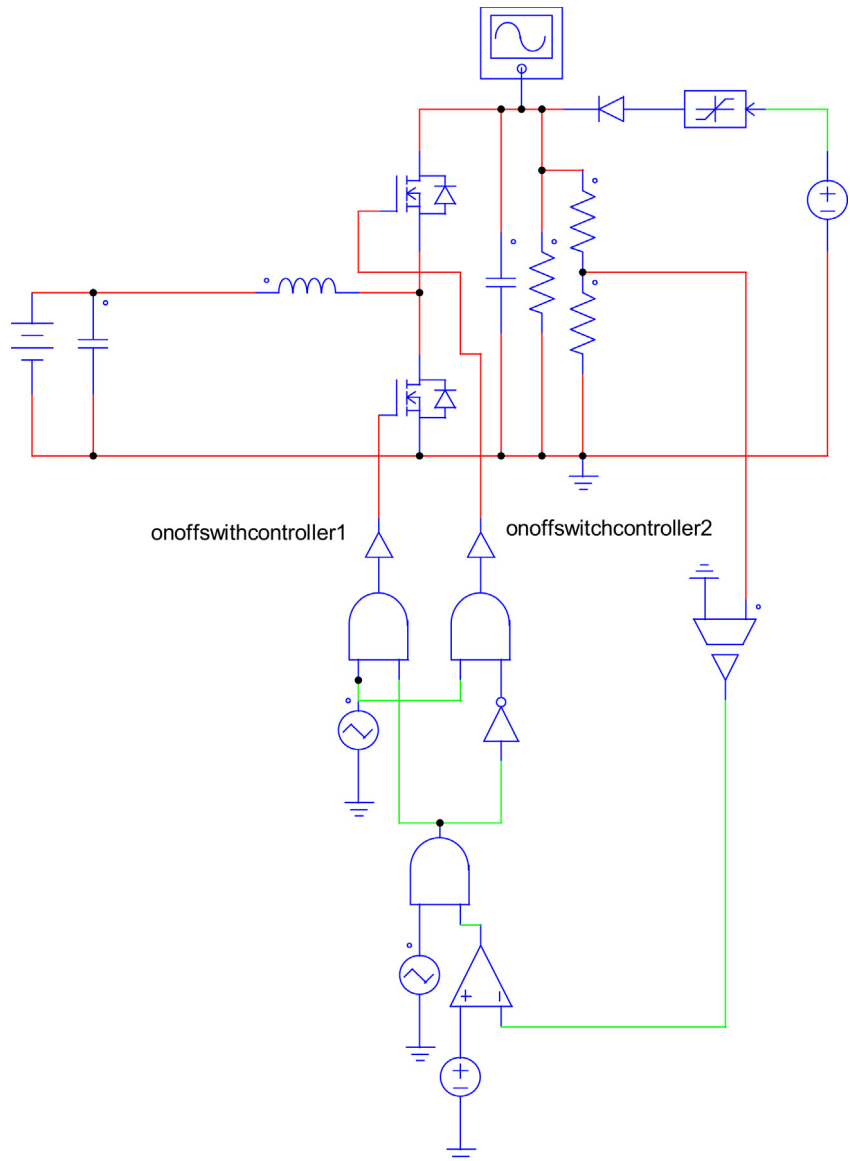

Fig. 5. Simulation of the bi-directional converter.

The bi-directional converter operates in two modes:

I. Buck mode: In this mode battery is charged through the bidirectional charger. Battery is charged by constant voltage, constant current methods. This procedure and simulation results are shown in Fig. 6.

I. **Buck mode.**

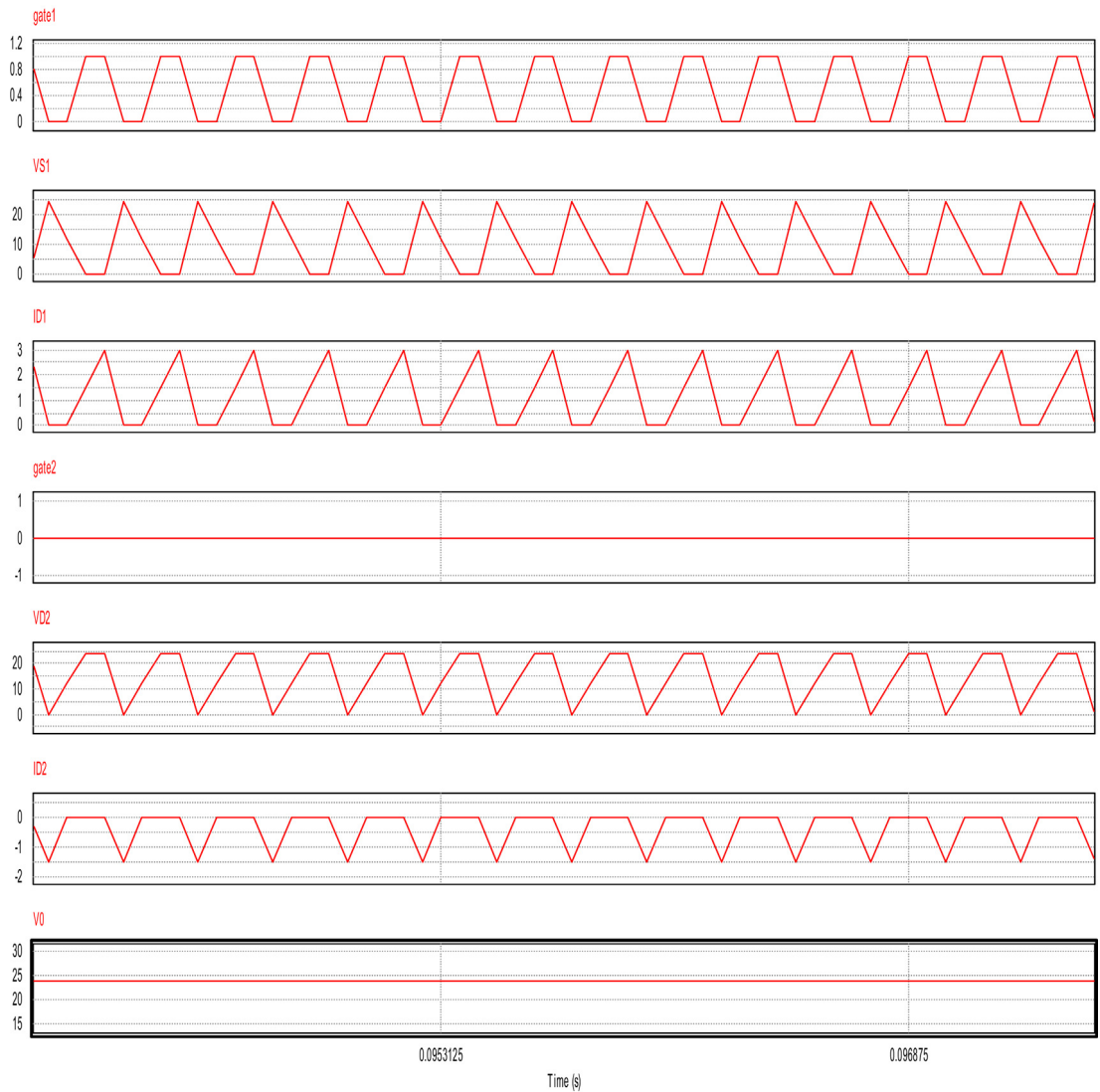

Fig. 6. Waveforms in buck mode.

II. Boost mode: In this mode battery is discharged through the bidirectional charger. Battery is discharged by constant voltage, constant current methods. This procedure and simulation results are shown in Fig. 7.

**II Boost Mode.**

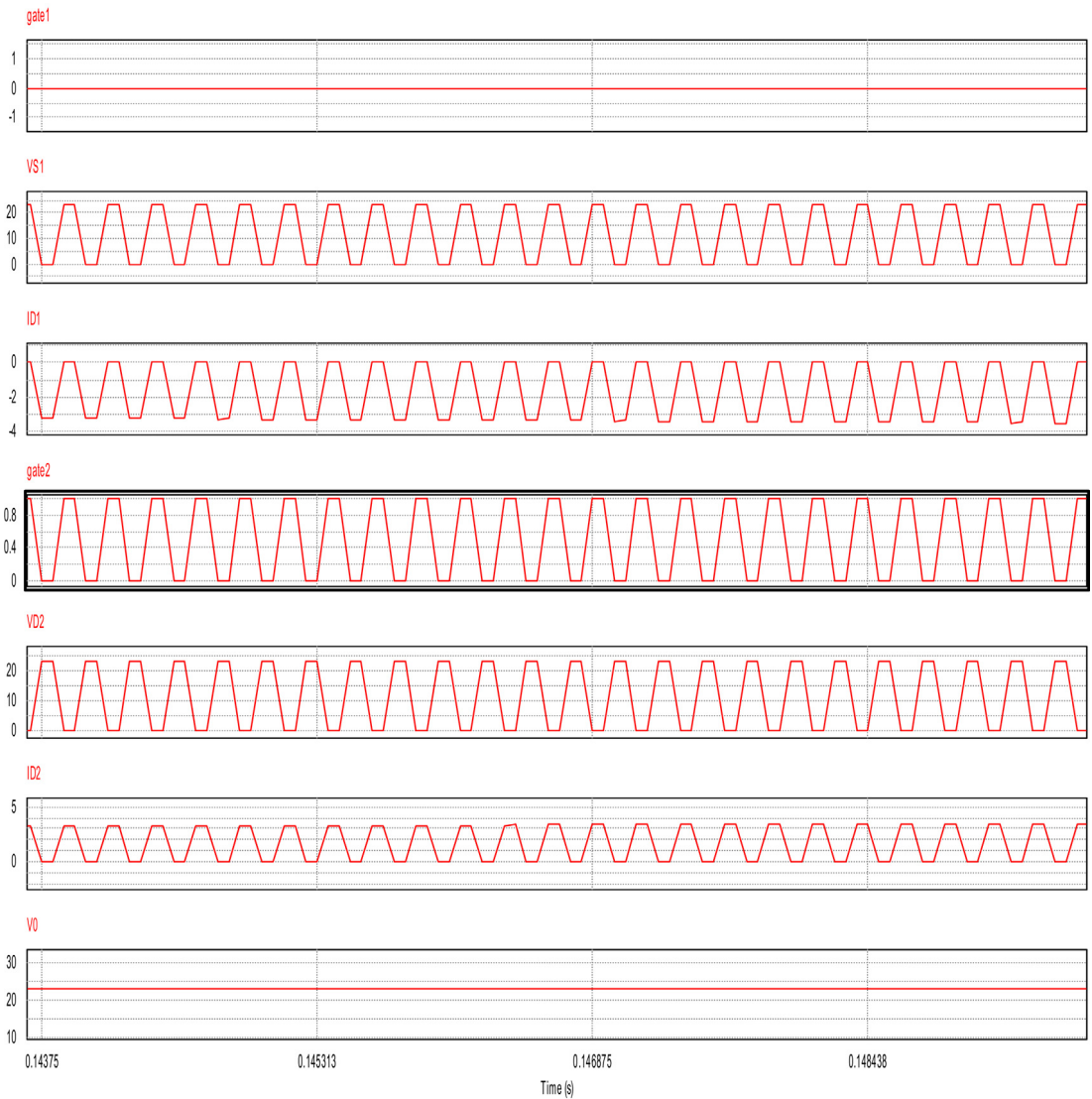

Fig. 7. Waveforms in boost mode.

## 6. Hardware implementation

### 6.1. Complete circuit diagram

See Fig. 8.

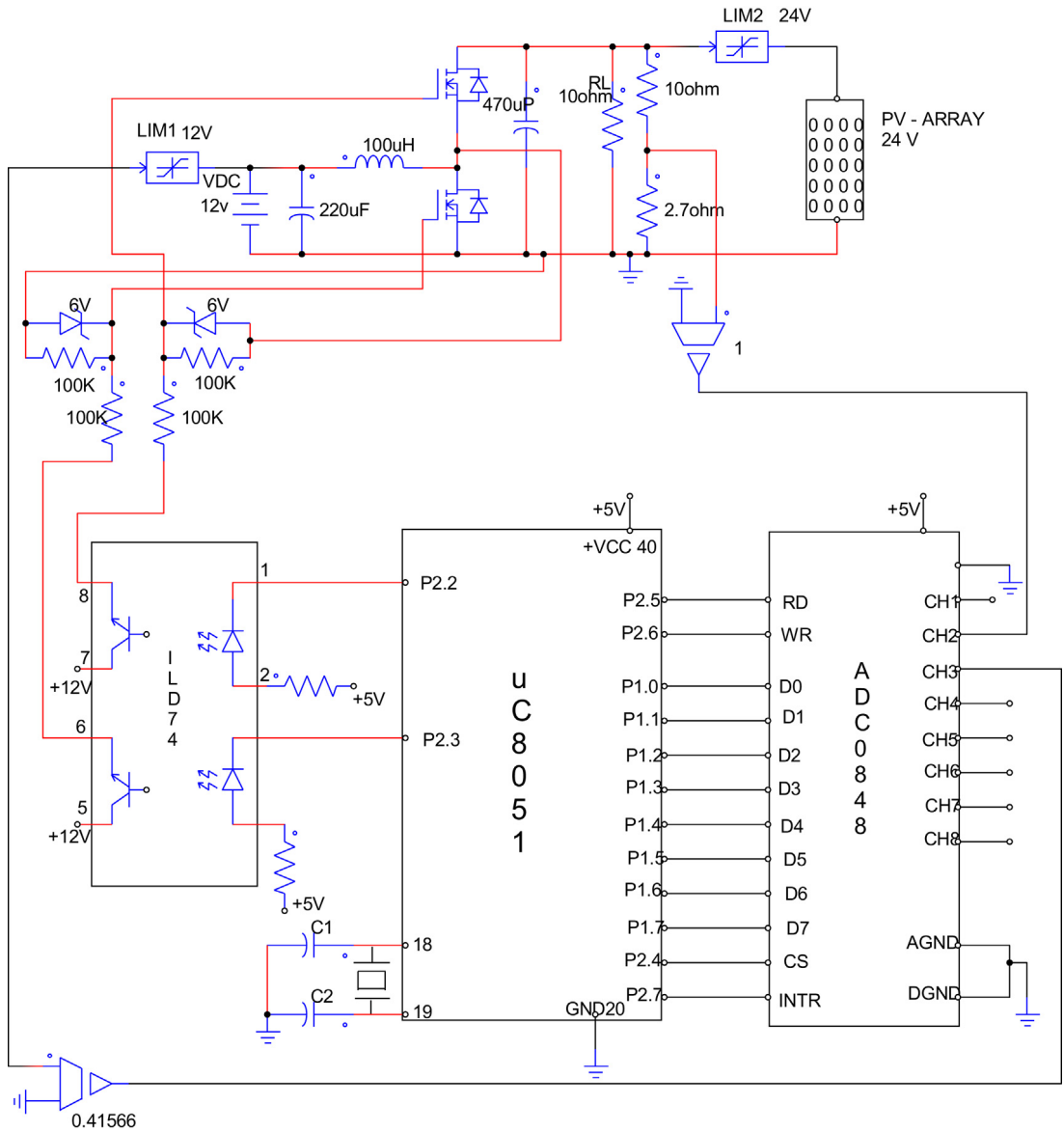

Fig. 8. Complete circuit diagram of microcontroller based bidirectional DC–DC buck–boost converter..

## *6.2. Flowchart*

See Fig. 9.

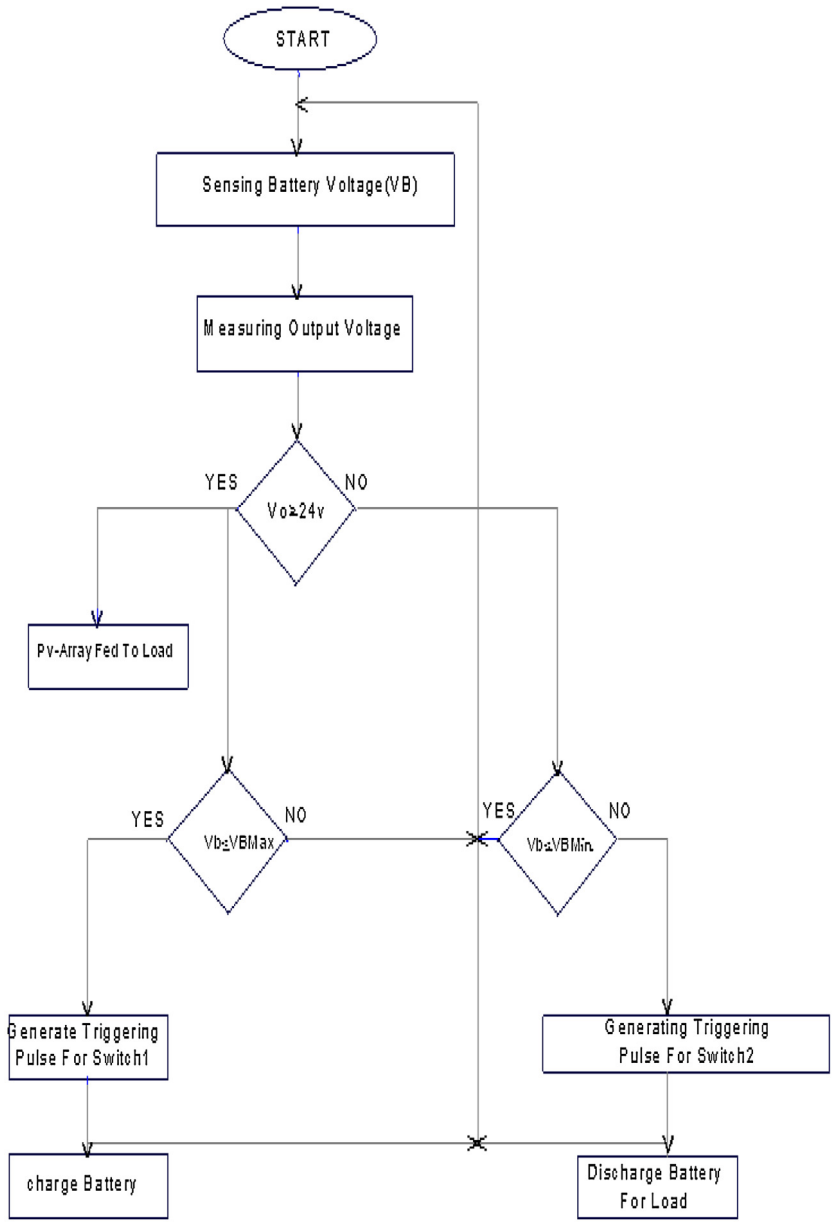

Fig. 9. Flowchart.

### 6.3. Hardware

See Figs. 10–12.

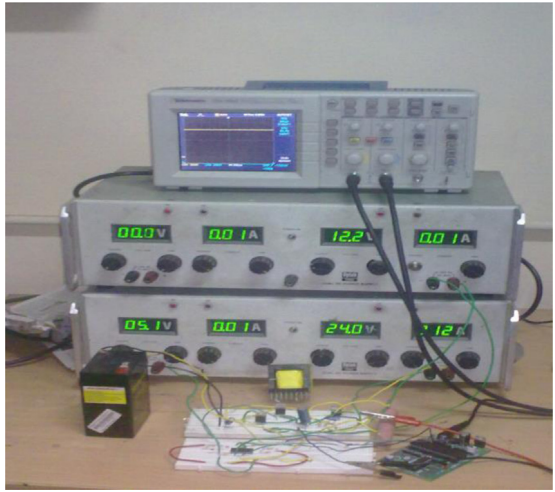

Fig. 10. Output voltage across a load.

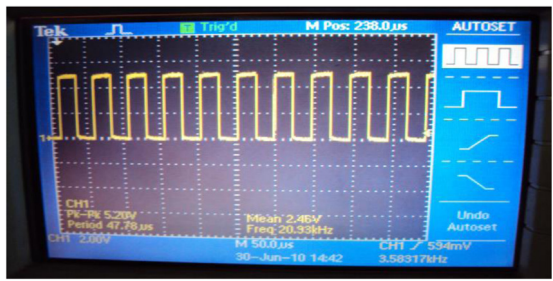

Fig. 11. Triggering pulse for switch s1.

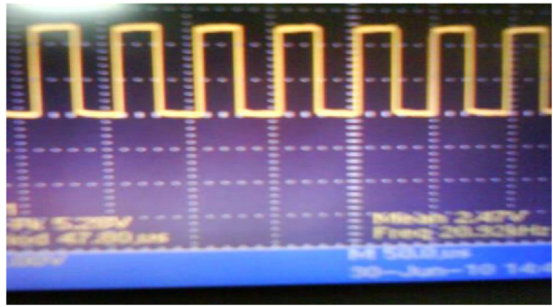

Fig. 12. Triggering pulse for switch s2.

### 6.4. Readings

**(I) Line regulation:**
See Table 1.

$$\begin{aligned}\text{Line regulation} &= \frac{\Delta Vout \times 100\%}{\Delta Vin} \\ &= \frac{0.003 \times 100\%}{5} \\ &= 0.06\%\end{aligned}$$

Table 1
Line regulation.

| Input voltage | Output voltage | Output current | Line regulation |
| --- | --- | --- | --- |
| 10 V | 23.902 V | +2.3902 A | 0.06% |
| 15 V | 23.905 V | +2.3905 A | |
| 20 V | 23.908 V | +2.3908 A | |
| 25 V | 23.997 V | +2.3997 A | |
| 39 V | 24.000 V | +2.4000 A | |

**(II) Load regulation:**
See Table 2.

$$\text{Load regulation} = \frac{[\text{Voltage(full load)} - \text{Voltage(Min load)}] \times 100\%}{\text{Voltage(Nominal load)}}$$
$$= \frac{[24.001 - 23.951] \times 100\%}{24}$$
$$= 0.208\%$$

Table 2
Load regulation.

| Load resistance | Output voltage | Output current | Load regulation |
| --- | --- | --- | --- |
| 6 Ω | 23.951 V | 3.991 A | 0.208% |
| 8 Ω | 23.952 V | 2.994 A | |
| 9 Ω | 23.955 V | 2.661 A | |
| 10 Ω | 23.957 V | 2.395 A | |
| 11 Ω | 23.959 V | 2.178 A | |
| 12 Ω | 24.001 V | 2.000 A | |

## 7. Conclusion

Charging controller has been designed for the battery charging and vehicle to grid concept has been demonstrated in the simulation results and hardware using microcontroller as well. It has to manage the charging and the discharging of the battery using microcontroller, and must therefore provide several flexible adjustment functions (e.g., load voltage, charging current) and by this arrangement the power from the battery is transfer to grid.